\documentclass[DIV=calc,paper=a4,fontsize=11pt,twocolumn]{scrartcl} 

\pdfoutput=1
\pdfmapfile{+txfonts.map}

\usepackage[english]{babel}
\usepackage[protrusion=true,expansion=true]{microtype}
\usepackage{amsmath,amsfonts,amsthm,amssymb}
\usepackage[final]{graphicx}
\usepackage{xcolor}
\usepackage[normal,small,hypcap,up,labelfont=bf,textfont=it]{caption}
\usepackage{subfig}
\usepackage{booktabs}
\usepackage{fix-cm}
\usepackage{dsfont}
\usepackage{bbm}
\usepackage{cite}
\usepackage[utf8]{inputenc}
\usepackage[perpage,symbol*]{footmisc}
\usepackage[varg]{txfonts}
\usepackage{fancyhdr}
\PassOptionsToPackage{hyphens}{url}\usepackage[pdfencoding=auto,psdextra]{hyperref}
\usepackage{bookmark}
\usepackage{verbatim}
\usepackage{fontenc}
\usepackage{cuted}

\usepackage{balance}

\DeclareCaptionFont{mycolor}{\color[HTML]{000000}}
\usepackage{sectsty}													
\allsectionsfont{
\color[HTML]{31ADF3}\usefont{OT1}{phv}{b}{n}
}

\sectionfont{
\color[HTML]{31ADF3}\usefont{OT1}{phv}{b}{n}
}

\usepackage{fancyhdr}												
\usepackage{lettrine}
\newcommand{\initial}[1]{%
\lettrine[lines=3,lhang=0.3,nindent=0em]{
\color[HTML]{31ADF3}
{\textsf{#1}}}{}}

\usepackage{titling}															

\newcommand{\HorRule}{\color[HTML]{31ADF3}
\rule{\linewidth}{1pt}%
}

\pretitle{\vspace{-30pt} \begin{flushleft} \HorRule
\fontsize{34}{34} \usefont{OT1}{phv}{b}{n} \color[HTML]{31ADF3} \selectfont
}
\title{The Enigmas of Fluctuations of the Universal Quantum Fields}					
\posttitle{\par\end{flushleft}\vskip 0.5em}

\preauthor{\begin{flushleft}\large \lineskip 0.5em \usefont{OT1}{phv}{b}{sl} \color[HTML]{31ADF3}}
\author{Mani L. Bhaumik\\[8pt]}											
\postauthor{\footnotesize \usefont{OT1}{phv}{m}{sl} \color[HTML]{000000}
Department of Physics and Astronomy, University of California, Los Angeles, USA. E-mail: \href{mailto:bhaumik@physics.ucla.edu}{bhaumik@physics.ucla.edu}\\[10pt]		
\scriptsize\usefont{OT1}{phv}{m}{n} \color[HTML]{31ADF3}{\textbf{Editors: \emph{Zvi Bern} \& \emph{Danko Georgiev}} }\\[5pt]
\color[HTML]{000000}{Article history: Submitted on November 25, 2023;  Accepted on December 8, 2023; Published on December 10, 2023.}
\par\end{flushleft}\HorRule}

\date{}																				

\begin{document}
\maketitle
\thispagestyle{fancy} 			
\initial{T}\textbf{he primary ingredients of reality are the universal quantum fields, which fluctuate persistently, spontaneously, and randomly. The general perception of the scientific community is that these quantum fluctuations are due to the uncertainty principle. Here, we present cogent arguments to show that the uncertainty principle is a consequence of the quantum fluctuations, but not their cause. This poses a conspicuous enigma as to how the universal fields remain immutable with an expectation value so accurate that it leads to experimental results, which are precise to one part in a trillion. We discuss some reasonable possibilities in the absence of a satisfactory solution to this enigma.\\ Quanta 2023; 12: 190--201.}

\begin{figure}[b!]
\rule{245 pt}{0.5 pt}\\[3pt]
\raisebox{-0.2\height}{\includegraphics[width=5mm]{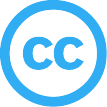}}\raisebox{-0.2\height}{\includegraphics[width=5mm]{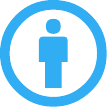}}
\footnotesize{This is an open access article distributed under the terms of the Creative Commons Attribution License \href{http://creativecommons.org/licenses/by/3.0/}{CC-BY-3.0}, which permits unrestricted use, distribution, and reproduction in any medium, provided the original author and source are credited.}
\end{figure}

\section{Introduction}

The ultimate ingredients of reality, unveiled so far by science, are
the abstract quantum fields that permeate all space of our unimaginably
vast universe for all times. All of material manifestations arise
from these universal quantum fields. Each elementary particle is a
quantum of its underlying field, which comprises the fundamental building
blocks of physical reality and are entrenched in our cherished Standard
Model of particle physics.\\

\noindent The esteemed Physics Nobel Laureate Steven Weinberg confirms:
\begin{quote}
The Standard Model provides a remarkably unified view of all types of
matter and force (except for gravitation) that we encounter in our
laboratories, in a set of equations that can fit on a single sheet
of paper. We can be certain that the Standard Model will appear as
at least an approximate feature of any better future theory. \cite{R1}
\end{quote}
Another distinguished Physics Nobel Laureate Frank Wilczek asserts:
``The standard model is very successful in describing reality---the
reality we find ourselves inhabiting'' \cite[p.~96]{R2}.
Wilczek additionally enumerates:
``The primary ingredient of physical reality,
from which all else is formed, fills all space and time. Every fragment,
each space-time element, has the same basic properties as every other
fragment. The primary ingredient of reality is alive with quantum
activity. Quantum activity has special characteristics. It is spontaneous
and unpredictable'' \cite[p.~74]{R2}.

These innately spontaneous and totally unpredictable activities of
the quantum fields are known as \emph{quantum fluctuations}. Thus,
unlike the stable classical fields, the quantum fields are distinctly
different in that they are incessantly teeming with intrinsic, spontaneous,
random activity all taking place locally in all space time elements,
from the infinitesimal to the infinite everywhere in this unimaginably
vast universe making it looking like an extremely busy place with
activities having infinite randomness.

These deeper properties of the quantum fields arise from the need
to introduce infinitely many degrees of freedom, and the possibility
that all these degrees of freedom are excited as quantum mechanical
fluctuations \cite[pp.~338-339]{R3}. ``Loosely speaking, energy can
be borrowed to make evanescent virtual particles. Each pair passes
away soon after it comes into being, but new pairs are constantly
boiling up, to establish an equilibrium distribution'' \cite[p.~404]{R3}.

Even though, we do not perceive its lively reality, indisputable evidence
of its existence can be found everywhere in nature with the help of
appropriate equipment because of the existence of the net equilibrium
distribution. Some of the distinct manifestations of the ubiquitous
quantum fluctuations will be presented following a brief discussion
of the discovery of the universal quantum fields.

\section{Discovery of the quantum fields}
\label{sec:2}

The existence of the quantum fields of nature came to light, totally
unexpectedly, from Paul Dirac's brilliant efforts \cite{R4} to combine
Schr\"{o}dinger's equation with special relativity, beginning in
1928. In addition to unveiling other important secrets of nature,
Dirac's continuing, arduous work eventually pointed in 1931 to the
possible existence of antiparticles like positrons with the same mass
and spin, but opposite charge of the electrons. Indeed, such a particle
was observed in August 1932 by Carl David Anderson \cite{R5} in the
cosmic ray tracks in a cloud chamber that led to his receiving the
Nobel Prize in Physics for 1936 \cite{Anderson1936}.

This discovery ultimately led to the concept of the underlying space
filling electron quantum field. When sufficient energy is provided,
the underlying quantum field would simultaneously create an electron-positron
pair. Such creation and annihilation of electron-positron pair was
copious in the early universe. However, because of some yet to be
completely understood leptogenesis process in the early universe \cite{R6},
a slight excess of leptons over anti-leptons were produced that left
a net excess of electrons that we observe today.

Nevertheless, the existence of an underlying universal electron quantum
field is established beyond any reasonable doubt because of the observed
fact that all electrons have exactly the same properties no matter
where and when they are produced, be in the early universe, in any
laboratory on earth today or even in the transient jets ejected in
astrophysical processes throughout the universe \cite{R7}. This phenomenon
appears to be true for all the particles listed in the Standard Model
\cite{R8} providing convincing existence of the underlying quantum
fields throughout the universe.

\section{Discovery of quantum fluctuations}
\label{sec:3}

The idea that empty space can have an intrinsic activity is seemingly
unintuitive. However, unlike the static classical fields, surprisingly
the quantum fields always fluctuate. The first indication of this
came from the father of the discovery of quanta, Max Planck himself.
After his exposition in 1900 that the energy of a single radiating
oscillator or a vibrating atomic unit comes in quanta, Planck proposed
\cite{R9} a new hypothesis for radiation. In a series of papers from
1911 to 1913, he reasonably established \cite{R10} that the energy oscillators
contained an additional term of $\frac{1}{2}\hbar\omega$, which marked
the birth of the concept of a \emph{zero-point energy}, as labeled
by Einstein. In 1916, Walter Nernst proposed \cite{R11} that even
empty space was filled with zero-point electromagnetic radiation.
Needless to say, the concept of this vacuum energy remained controversial.
Even Einstein once proclaimed that the zero-point energy is dead as
a doornail. However, this perception changed in 1924, when Robert
Mulliken \cite{R12} provided direct evidence. Using the band spectrum
of the boron monoxide isotopes \textsuperscript{10}BO and \textsuperscript{11}BO,
he showed that in contrast to the observed spectra, the transition
frequencies between the ground vibrational states would disappear
if there was no vacuum energy.

With the advent of the quantum mechanics, by the summer of 1926 the
zero-point energy was no longer controversial, at least not in so
far as it concerned material systems. Calculation \cite[pp.~155-156]{R13}
using the Schr\"{o}dinger equation shows that the ground state of
a quantum harmonic oscillator has the minimum energy of $\frac{1}{2}\hbar\omega$,
which is attributable solely to the zero-point energy of vacuum fluctuation.
Because of these vacuum fluctuations alone, the \emph{standard deviations},
$\sigma_{x}$ and $\sigma_{p}$, of the Fourier-conjugated Gaussian
position and momentum wave functions of the ground state of a quantum
harmonic oscillator cannot become zero but satisfy the relation $\sigma_{x}\cdot\sigma_{p}=\frac{1}{2}\hbar$.
Thus, the uncertainty relation is not sufficient for quantum fluctuations
and hence they should be spontaneous.

\section{Fluctuations of the radiation fields}

Despite the origin of the concept of a quantum in the theory of thermal
radiation, quantum mechanics in its early stages remarkably dealt
only with material particles and not with radiation itself. The initial
application of the new quantum theory to fields rather than particles
came in 1926 in a paper by Born, Heisenberg and Jordan \cite{R14}.
By applying the same mathematical techniques used for material oscillator,
they were able to show that the energy of each mode of oscillation
of an electromagnetic field was quantized with the basic unit of $\hbar\omega$.
But they dealt only with the EM radiation in empty space without interaction
with any matter and as such their efforts did not lead to any significant
prediction.

Paul Dirac's fundamental paper in 1927 \cite{R15} radically changed
the situation. In the presence of atoms or other system of charged
particles, Dirac calculated the interaction energy between the field
and an atom and used it as a perturbation upon the energy of the atom
leading to some spectacular results. However, nature was still conceived
to be composed of two very different ingredients of \emph{particles}
and \emph{fields} that needed description in terms of quantum mechanics
but in quite different ways. Again, as described earlier in Section
\ref{sec:2}, the existence of the all-pervading quantum fields of
nature came to light from Dirac's brilliant efforts \cite{R4} to
combine Schr\"{o}dinger's equation with special relativity, beginning
in 1928.

In spite of all the outstanding successes of these efforts, some serious
difficulty turned up and it took quite a while to resolve it. The
problem essentially arises from the existence of the spontaneous vacuum
fluctuations without application of any energy. These vacuum fluctuations
produce copious amount of virtual particle and antiparticle pairs
that significantly creates screening of the intrinsic mass and charge
of a particle to provide their measured values. The self-energy of
the electron, especially considering the higher-order perturbative
calculations in quantum electrodynamics (QED) always turned out to
be infinite as was first pointed out by J. R. Oppenheimer in 1930
\cite{R16}.

An infinite self-energy appears not only when the electron is moving
in an orbit, but also when it is at rest in empty space. Therefore,
the electron mass and charge listed in tables of physical data could
not be just the \emph{bare mass} and \emph{bare charge}, the quantities
that appear in our equations for the electron field, but should be
identified with the bare mass and charge together with the infinite
self-mass and self-charge, produced by the interaction of the electron
with its own virtual cloud. Eliminating the infinities by a redefinition
of physical parameters has come to be called \emph{renormalization}.
Although Dirac and others were totally against such a procedure, no
inconsistency arises since we can never turn off the electron's virtual
photon cloud because of the ubiquitous spontaneous fluctuations of
the quantum fields.

\section{Effects due to quantum fluctuations}

\subsection{Lamb shift}

In the historic Shelter Island conference in 1947, perhaps the most
exciting report was presented by Willis Lamb. Using the great advances
in microwave technology developed during the war, Lamb and Retherford
\cite{R17} presented convincing evidence showing that the two supposedly
degenerate energy levels $^{2}S_{\frac{1}{2}}$ and $^{2}P_{\frac{1}{2}}$
of the hydrogen atom are separated by 1057.862 MHz compared to the
calculated value of 1057.864 MHz.

This was not predicted by the Dirac equation, according to which these
states should have the same energy. The observed shift now known as
the Lamb Shift was named after Willis Lamb that led to the essential
validation of renormalization. Shortly after its discovery, Hans Bethe
\cite{R18} was the first one to derive the Lamb shift by implementing
the idea of mass renormalization, which allowed him to present a somewhat
preliminary calculation, without relativity, in support of the observed
energy shift.

Eventually, a very erudite group of physicists including Julian Schwinger,
Shin-Ichiro Tomonaga, Richard Feynman and Freeman Dyson \cite{R19}
developed a reliable way by incorporating infinite ``counter terms''
in the Hamiltonian to compensate for the infinite mass and charge.
The Lamb shift currently provides a measurement of the fine-structure
constant $\alpha$ to better than one part in a billion \cite{R20},
allowing a precision test of quantum electrodynamics and a robust
testimonial to the existence of ubiquitous quantum fluctuations.

\subsection{Anomalous electron g-factor}

Another subject that drew attention at the Shelter island conference
is the anomalous $g$-factor of the electron spin. The spin magnetic
moment of the electron was derived by Dirac \cite{R4,R21} in his
pivotal work in 1928 to have a value of 2. In advanced QED, the anomalous
magnetic moment of the electron spin arises due to the effect of quantum
fluctuations as in Lamb Shift. The value was first measured by Polykarp
Kusch and Henry M. Foley \cite{R22} shortly before the conference.
A more accurate value of $\frac{g}{2}=1.00119$ was presented by Foley
and Kusch \cite{R23,R24}. Detailed discussion of the anomalous magnetic
moment has been provided by Kusch in his Nobel Lecture \cite{Kusch1955,R25}. Julian Schwinger \cite{R26}
was the first to present a theoretical derivation of the anomalous
$g$-factor caused by vacuum fluctuations to be $\frac{g}{2}=1.001162$.

The latest measurement of the anomalous magnetic dipole moment of
the electron spin is provided by Fan et al. \cite{R27} to be $\frac{g}{2}=1.00115965218059(13)$,
a precision better than one part in a trillion. This represents one
of the most accurate measurements in all of physics, again providing
a glorious testimonial to the existence of quantum fluctuations.

Willis Lamb and Polycarp Kusch shared the 1955 Nobel Prize in Physics
for their work that crucially depended upon the existence of quantum
fluctuations \cite{Lamb1955,Lamb1956,Kusch1955,R25}.

\subsection{The Casimir effect}

As predicted by Hendrik Casimir \cite{R28} in 1948, there is an attractive
force created between two uncharged perfectly parallel metal plates
inserted in a quantum vacuum. This effect occurs because the plates
would be slightly pushed towards each other as some of the quantum
fluctuations of the electromagnetic quantum field, with wavelength
longer than the distance between the plates, would not be sustainable
in the space between the metal plates. Although simple in principle,
the actual experiment turned out to be quite difficult.

\begin{figure}
\begin{centering}
\includegraphics[width=85mm]{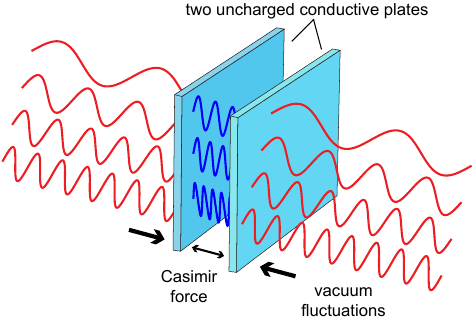}
\par\end{centering}

\caption{\label{fig:1}A schematic outline for direct demonstration of the
source of a physical force in the Casimir effect arising from quantum
fluctuations of the electromagnetic quantum field.}
\end{figure}

With a very carefully designed experiment, the effect of the small
force was conclusively demonstrated by Steve Lamoreaux \cite{R29}
in 1997. This is considered as a simple but direct proof of the existence
of quantum fluctuations in the vacuum as illustrated in Fig. \ref{fig:1}.

\subsection{Spontaneous emission of radiation}

According to the Schr\"{o}dinger equation, any stationary excited
state by itself should have an infinite lifetime if nothing disturbs
it. With no light in the universe whatsoever, it would be hard to
imagine the existence of intelligent life in it! However, Dirac's
1927 masterpiece \cite{R15}, revealed that the \emph{spontaneous
emission} of radiation from an excited state of an atom or a molecule
appears as a forced emission caused by the vacuum fluctuations of
the electromagnetic field.

Thus, to explain spontaneous transitions, quantum mechanics must be
extended, whereby the electromagnetic field is quantized at every
point in space. The quantum field theory of electrons and electromagnetic
fields was labeled by Dirac as \emph{quantum electrodynamics} (QED).
In QED, the spontaneously emitted photon has infinite different modes
to propagate into, thus the probability of the atom re-absorbing the
photon and returning to the original state is negligible, making the
atomic decay practically irreversible. If one were to keep track of
all the vacuum modes, the combined atom-vacuum system would undergo
unitary time evolution, making the decay process reversible.

However, in cavity QED the decay rate, transition energy, and radiation
pattern of spontaneous emission can all be altered by modifying the
vacuum field fluctuations by a cavity wall. The coupling of atom and
vacuum fields was first formulated by Jaynes and Cummings \cite{R30}
in 1963, in which it was predicted that spontaneous emission becomes
even reversible if an atom is put in a high-Q single-mode cavity.

\begin{figure}
\begin{centering}
\includegraphics[width=85mm]{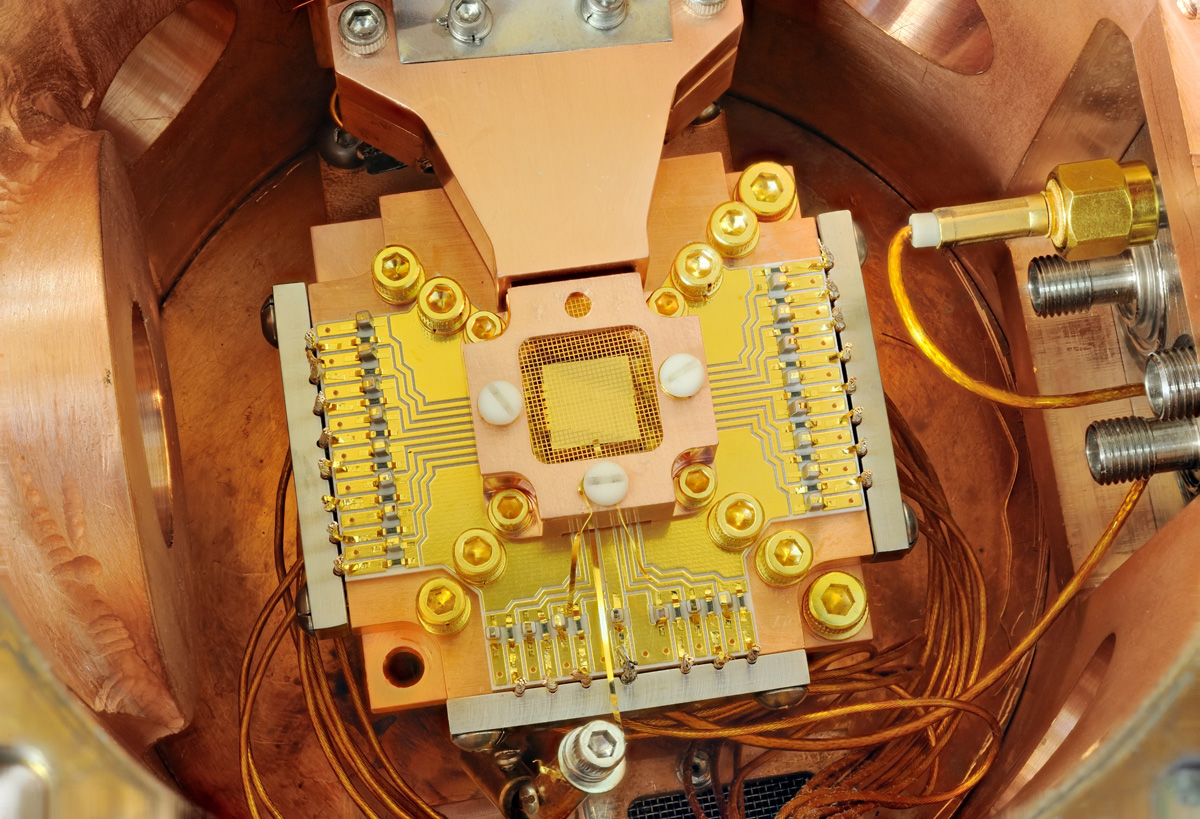}
\par\end{centering}

\caption{\label{fig:2}Chip ion trap for quantum computing at NIST.
Credit:~\href{https://www.nist.gov/image/quantumcomputingiontrappingjpg}{Y. Colombe, National Institute
of Standards and Technology, U.S. Department of Commerce}.}
\end{figure}

In recent times, the subject of cavity QED has advanced immensely.
The 2012 Nobel Prize for Physics was awarded to Serge Haroche and
David Wineland for their work on controlling such quantum systems
\cite{Haroche2012,Haroche2013,Wineland2012,Wineland2013} and thus attesting yet again the
existence and importance of vacuum quantum fluctuations. The techniques
developed to create and measure cavity QED states are now being applied
to the development of quantum computers. An example is shown in Fig.
\ref{fig:2}.

\subsection{Vacuum polarization}

\begin{figure}
\begin{centering}
\includegraphics[width=76mm]{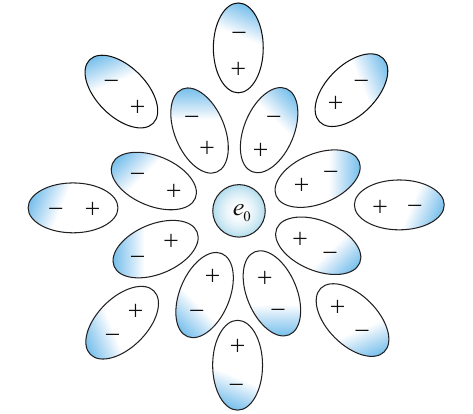}
\par\end{centering}

\caption{\label{fig:3}Schematic snapshot of screening of the bare electron
charge $e_{0}$ is performed by virtual electron-positron pairs, which
are effective dipoles as discussed in Ref.~\cite{R7}.}
\end{figure}

In quantum field theory (QFT), the bare mass or charge of an elementary
particle like electron is the ultimate upper limit of its mass or
charge, which is presumed to be infinite. It differs from the measured
mass or charge because the latter includes the \emph{screening} of
the particle by pairs of virtual particles that are temporarily created
by the quantum fluctuation around the particle. This is depicted in
Fig. \ref{fig:3} and is known as \emph{vacuum polarization}. At smaller
distances as we begin to penetrate the polarization cloud, we come
closer to the bare charge or mass. The range of the correction term
is roughly the Compton wave length. Usually, the bare mass and bare
charge are included in the Lagrangian while only the physical mass
and charge are taken as observables. This is known as \emph{renormalization},
which had played an essential role in the Standard Model.

\subsection{Running of coupling constants}

As mentioned earlier, the bare charge is \emph{infinite} as is the bare mass.
The effective charge and effective mass change with the
energy scale as we penetrate the screening cloud and are determined
by the local surrounding screening of charge and mass by virtual particles.
While virtual particles obey conservation of energy and momentum,
they can have any energy and momentum, even one that is not allowed
by the relativistic energy--momentum relation for the observed mass
of that particle. Such a particle is called off-shell.

In QFT, a coupling constant or gauge coupling parameter is a number
that determines the strength of the force exerted in an interaction.
The coupling constants are not really constant as they depend on the
energy scale. One can change the energy scale and thus observe different
values for the coupling constant as one penetrates the surrounding
shielding created by vacuum fluctuations of appropriate fields with
higher energy. A special role is played in relativistic quantum theories
by couplings that are dimensionless, i.e., are pure numbers. An example
of a dimensionless constant is the fine-structure constant $\alpha$.

\begin{figure}
\begin{centering}
\includegraphics[width=85mm]{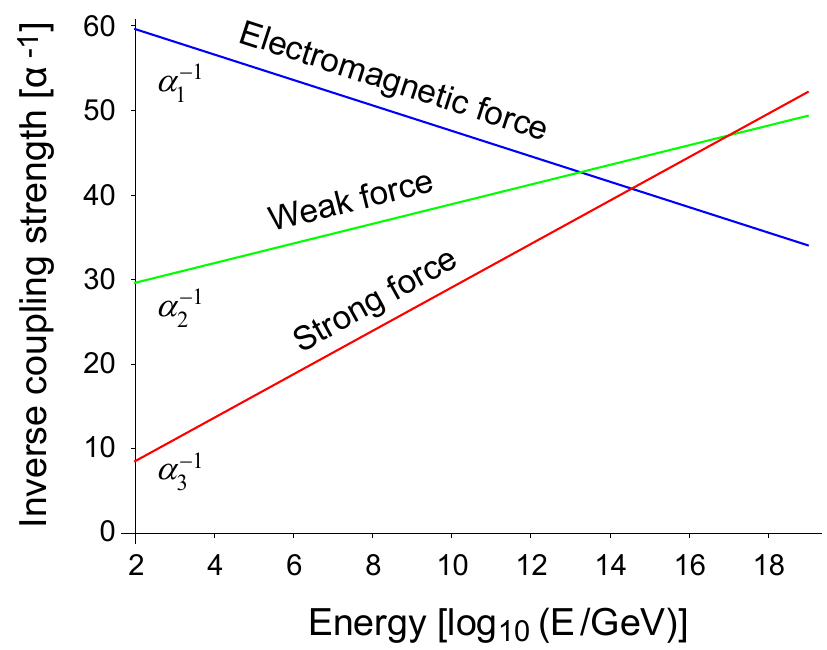}
\par\end{centering}

\caption{\label{fig:4}The inverse of the running coupling constants $\alpha_{1}^{-1}$,
$\alpha_{2}^{-1}$and $\alpha_{3}^{-1}$, respectively for the electromagnetic, weak and strong force, as a function of energy.
The explicit formulas for the running coupling constants can be found
in Ref.~\cite{Lautsch2014}.}
\end{figure}

The value of the coupling constant is said to \emph{run} with energy,
and the constants themselves are usually referred to as running coupling
constants. The inverse of the running coupling constants of the electromagnetic, weak and strong force are plotted in Fig. \ref{fig:4}.
This is another testament of the existence of the vacuum fluctuations of three different force fields.

\subsection{Intermingling of quantum fields}

\begin{figure*}[t!]
\begin{centering}
\includegraphics[width=140mm]{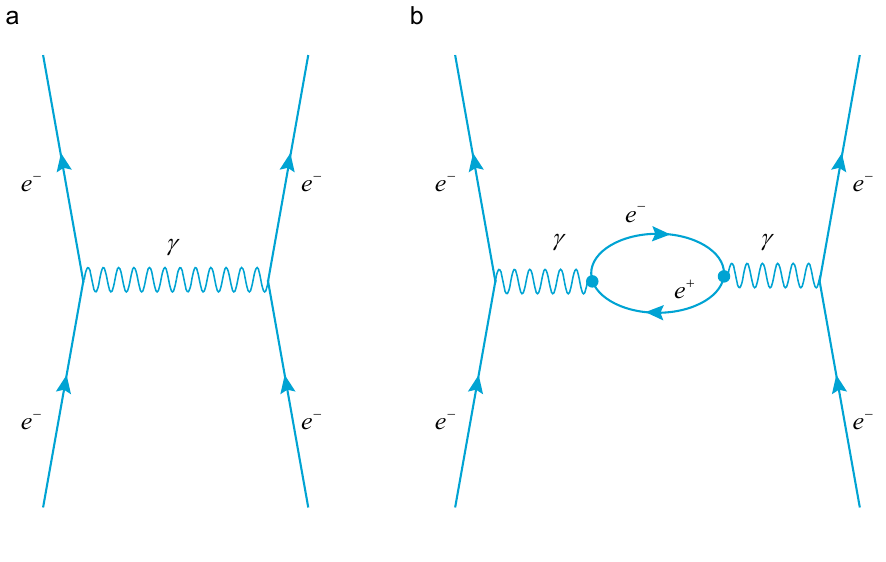}
\par\end{centering}

\caption{\label{fig:5}The electromagnetic force between electrically charged particles.
(a) Exchange of a virtual photon $\gamma$ between two electrons $e^{-}$.
(b) The electromagnetic field can be affected by spontaneous activity in the electron field through transient creation of electron-positron pairs, $e^{-} + e^{+}$, as discussed in Ref.~\cite[p.~90]{R2}. In the Feynman diagrams, time flows vertically such that the bottom of the diagram is the past and the top is the future, whereas a single spatial dimension is depicted horizontally.}
\end{figure*}

Each of the quantum fields whose respective particles are ensconced
in the Standard Model of particle physics contains a small amount
of all the other fields. This is primarily due to the fact that a
quantum fluctuation, which is an irregular disturbance of a field
causes similar disturbances in all the other fields in succession
(Fig.~\ref{fig:5}).

Quoting Frank Wilczek:
``The electromagnetic field gets modified by its interaction with a spontaneous fluctuation
in the electron field---or, in other words, by its interaction with
a virtual electron-positron pair. {[}$\ldots${]} They lead to complicated,
small but very specific modifications of the force you would calculate
from Maxwell's equations. Those modifications have been observed,
precisely, in accurate experiments'' \cite[p.~89]{R2}.
Emphasizing Wilczek's critical observation again that despite the precipitous transitory characteristics
of the virtual particles, there is a residual equilibrium distribution \cite[p.~404]{R3}.

So, we ascertain that particles arising out of the quantum fields
are not just simple objects, and although sometimes people naively describe
them as simple ripples in a single field, that is far from true. Only
in a universe with no spontaneous activities---with no interactions
among quantum fields at all---are particles merely ripple in a single field!
In fact, we know quite explicitly what the fields are, out of which
a physical particle is built, at least order by order in perturbation
theory.

The irregular disturbances of the fields relate to virtual particles
since their respective energy-momentum does not correspond to the
physical mass of a particle. One says that these particles are off-shell.
However, in the process, the total energy-momentum is exactly conserved
at all times. Because of the self-interaction of the quantum fields,
such an energy-momentum eigenstate of the field can be expressed as
a specific Lorentz covariant superposition of field shapes of the
electron field along with all the other quantum fields of the Standard
Model of particle physics. It is particularly important to emphasize
again that the quantum fluctuations are transitory but new ones are
constantly boiling up to establish an equilibrium distribution so
stable that their contribution to the screening of the bare charge
provide the measured charge of the electron to be accurate up to nine
decimal places \cite{R31} and the electron $g$-factor results in
a measurement accuracy of better than a part in a trillion \cite{R27}.

\subsection{Seeding of galaxies from quantum fluctuations}

\begin{figure*}
\begin{centering}
\includegraphics[width=169mm]{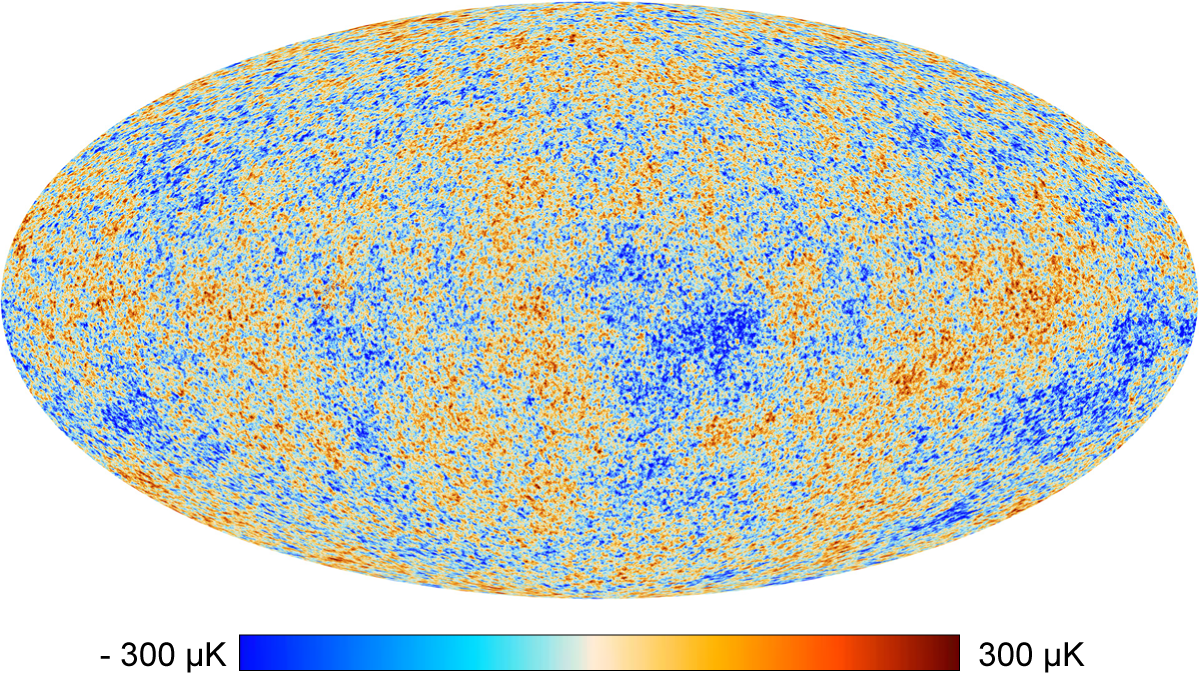}
\par\end{centering}

\caption{\label{fig:6}Cosmic microwave background radiation observed by the
Planck satellite. This is a very graphic demonstration of colossally
enlarged universal quantum fluctuations. Credit: \href{https://www.esa.int/ESA\_Multimedia/Images/2013/03/Planck\_CMB}{ESA and the Planck
Collaboration}.}
\end{figure*}

The interpretation of the observed temperature anisotropies in the
Cosmic Microwave Background (CMB) shown in Fig. \ref{fig:6} is the
result of density perturbations which seeded the formation of the
large-scale structures of galaxies, clusters, and superclusters that
we observe today. The discovery of temperature anisotropies by COBE,
WMAP, and Planck satellites provides evidence that such density inhomogeneities
existed in the early universe, most likely caused by quantum fluctuations
in the scalar inflaton field \cite{R32}. Since it requires a lot
more energy for the primordial photons to overcome the gravitational
pull and exit the denser potential wells, these areas actually end
up having less energy and are colder as shown in blue speckles than
the less dense regions of higher energy shown in red.

When we look at the CMB radiation, we are looking at roughly $40000$ causally
disconnected regions of the universe. How is it, then, that each of
these has the same temperature to one part in $10^{5}$? Furthermore,
if large-scale structure grew via gravitational infall from tiny inhomogeneities
in the early universe, where did these primordial perturbations come
from? Such perturbations are produced naturally during inflation,
a period of exponential expansion in the early universe that makes
it simpler and smoother.

\pagebreak
In 1981, a young post-doctoral fellow, Alan Guth at the Stanford University
presented a visionary article \cite{R33} that revolutionized our
concept about the very early universe. He persuasively pointed out
that to satisfactorily explain some of the mysterious aspects such
as the homogeneity, flatness, etc., of the early universe, there had
to be an extraordinarily fast expansion of the space itself by a factor
of nearly $10^{30}$ by a process he dubbed \emph{inflation}. It was
presumably brought about by the scalar inflaton quantum field, details
of which are yet to be clearly understood.

The 2014 Kavli Prize in Astrophysics was awarded to Guth for his pioneering
contributions to the theory of cosmic inflation. The predictions of
the simplest versions of the theory have been so successful that most
cosmologists accept that some form of inflation truly did occur in
the very early universe. Assuming the veracity of cosmic inflation
at the dawn of the universe, the imprint of the disturbances of the
inflaton quantum field could manifestly be discernible as immensely
enlarged vacuum fluctuations of the field in the cosmic microwave
background radiation anisotropy, as described earlier, both by the
WMAP and Planck satellite \cite{R34}. Thus, a very graphic demonstration
of the existence of the fluctuations of a cosmic quantum field is
cogently demonstrated.

In fact, quantum fluctuations are more omnipresent than we perceive.
Each fundamental particle is a quantum of its respective underlying
quantum field. As pointed out earlier, each of the quantum fields
contains a small amount of all the other fields due to mutual interaction
of the fields caused by spontaneous vacuum fluctuations. Thus, every
elementary particle is a witness to the universal spontaneous quantum
fluctuations. How much more ubiquitous anything can be?

\section{Fluctuation of quantum fields and uncertainty principle}

Despite the overabundance of evidences listed above, demonstrating
that the fluctuations of the universal quantum fields are inherently
spontaneous, there seems to be a pervasive view in much of the scientific
community that quantum fluctuations are caused by the uncertainty
principle. This could perhaps be attributed to the fact that the importance
of quantum fluctuations was not properly appreciated until 1970s \cite[p.~124]{R35}
while the idea of the uncertainty principle originated during the
very early days of quantum mechanics in 1927. Consequently, the uncertainty
principle also gloriously known as the ``famous Heisenberg uncertainty
principle,'' with its somewhat of a volatile history, has become almost
a slogan in quantum physics.

A conspicuous example can be found in a review \cite{R36} by
the famed physicist Victor Weisskopf where he states:
\begin{quote}
In quantum mechanics an oscillator cannot be exactly at its rest
position except at the expense of an infinite momentum, according
to Heisenberg's uncertainty relation. \cite[p.~71]{R36}
\end{quote}
The contention that an ``oscillator cannot be exactly at its rest
position'' to avoid an ``infinite momentum'' is rather misleading.
In Section \ref{sec:3}, we have confirmed that the ground state of
a quantum harmonic oscillator has the minimum energy of $\frac{1}{2}\hbar\omega$,
which is attributable solely to the zero-point energy of vacuum fluctuations.

The uncertainty relation is merely a relation between the standard
deviations, $\sigma_{x}$ and $\sigma_{k}$, of position and wavenumber
wave functions, which are Fourier transforms of each other. For a
Gaussian wave packet with \emph{standard deviations}, $\sigma_{x}$
and $\sigma_{k}$, respectively computed in position and wavenumber
basis, the minimum uncertainty relation is obeyed
\begin{equation}
\sigma_{x}\cdot\sigma_{k}=\frac{1}{2}
\end{equation}
since for a Gaussian, $\sigma_{x}=\frac{1}{2\sigma_{k}}$ .

In modern quantum mechanical literature, it is common to define \emph{standard
deviations} using $\Delta$ symbol and expectation values \cite{R37}
\begin{align}
\Delta x & \equiv\sigma_{x}=\sqrt{ \left\langle \hat{x}^{2}\right\rangle -\left\langle \hat{x}\right\rangle {}^{2} }\\
\Delta k & \equiv\sigma_{k}=\sqrt{ \left\langle \hat{k}^{2}\right\rangle -\left\langle \hat{k}\right\rangle {}^{2} }
\end{align}

Following the original formulaion of the uncertainty principle by
Heisenberg \cite{R38,R39}, the \emph{standard deviations} $\Delta x$
and $\Delta k$ were interpreted by Kennard \cite{R40} as ``measures
of indetermination'' of the corresponding physical observables. The
mathematical form of the minimum uncertainty relation is not affected
by the interpretation of the standard deviations as physical uncertainties
\begin{equation}
\Delta x\cdot\Delta k=\frac{1}{2}
\end{equation}
Such an uncertainty relation is valid for both classical and quantum
mechanical functions. However, it becomes very significant in quantum
mechanics as we introduce, for example, the quantum mechanical observation
of Louis de Broglie of a matter wave with length $\lambda=h/p$, or
equivalently $p=\hbar k$, leading to the quantum uncertainty principle
between position $\hat{x}$ and momentum $\hat{p}$ observables to
be
\begin{equation}
\Delta x\cdot\Delta p=\frac{1}{2}\hbar
\end{equation}
Thus, the uncertainty relation between position and momentum follows
from the simple quantum mechanical relationship $p=\hbar k$ or more
meticulously from the Hamiltonian of the quantum harmonic oscillator,
which governs the temporal dynamics (quantum fluctuations) of the
ground state (see also the \hyperref[sec:9]{Appendix}). It is a consequence of the spontaneous
quantum fluctuations of the ground state that the position and momentum
observables obey the uncertainty relation, but it is not the other
way around. Contrary to the apparent general notion, the uncertainty
principle by itself cannot cause anything to fluctuate. It merely
gives the relationship of an effect and has nothing to do with the
cause.

\section{History of the uncertainty principle}

The somewhat of a checkered history of the uncertainty principle started
with the publication \cite{R38,R39} of Werner Heisenberg in 1927.

During this time Heisenberg was working in Copenhagen with the legendary
Niels Bohr. In spite of a request from Bohr, who was on a vacation
at the time, to hold off submission of the paper, Heisenberg went
ahead and sent it for publication, immensely disturbing Bohr. This
resulted in Heisenberg's submission of an ``Addition in Proof'' to
the paper to correct for some valid objections of Bohr.

Using a thought experiment with a $\gamma$ ray microscope, Heisenberg
argued that the product in the \emph{noise} in a position measurement
and the momentum \emph{disturbance} caused by that measurement should
be no less than $\frac{1}{2}\hbar$. Masanao Ozawa \cite{R41} pointed
out the shortcoming of Heisenberg's derivation.

Shortly after Heisenberg's original publication, Earle Kennard \cite{R42}
provided the correct relationship relating the \emph{standard deviation}
of position $\sigma_{x}$ and the standard deviation of momentum $\sigma_{p}$
for any quantum wavefunction
\begin{equation}
\sigma_{x}\cdot\sigma_{p}\geq\frac{1}{2}\hbar
\end{equation}

Heisenberg attempted to show that this relation is a straightforward
mathematical consequence of the commutation relationship derived by
Born and Jordan \cite{R43}
\begin{equation}
\hat{p}\hat{x}-\hat{x}\hat{p}=-\imath\hbar
\end{equation}
Based on the above equation, Heisenberg proposed an energy-time commutation
relationship
\begin{equation}
\hat{E}\hat{t}-\hat{t}\hat{E}=-\imath\hbar
\end{equation}
which would lead to
\begin{equation}
\Delta E\cdot\Delta t\geq\frac{1}{2}\hbar
\end{equation}
According to Paul Busch, the time--energy relationship has been a controversial
issue since its beginning \cite{R44}. He ascertains:
\begin{quote}
Different types of time--energy uncertainty relation can indeed be deduced in
specific contexts, but there is no unique universal relation that
could stand on equal footing with the position--momentum uncertainty
relation. \cite[p.~69]{R44}
\end{quote}
In conclusion of his presentation, Busch summarizes the main types
of time--energy uncertainty relations and their range of validity depending
on the physical interpretation of the quantities $\Delta E$ and $\Delta t$.
In the case of vacuum fluctuations, it would be reasonable to derive
an energy--time uncertainty relationship from the relativistic relation
$E=h\nu$ since it treats space and time on an equal footing, as we
would expect for a relativistic theory like QFT. To facilitate conservation
of energy, we can then use the relationship $ E\cdot t=h$
for quantum fluctuations, where $ t=1/\nu$.

In the absence of even a valid universal energy--time uncertainty
relationship, how could we then even think that the quantum fluctuations
are caused by the uncertainty principle? It can merely be a relationship
between the effects and not the cause. One can only anticipate that
the paradigm is changing. In fact, in a popular modern textbook \cite{R35},
we are starting to notice statements like the following:\\

``Incidentally, the vacuum in quantum field theory is a stormy sea
of quantum fluctuations.'' \cite[p.~20]{R35}\\

``The fluctuating quantum field is real.'' \cite[p.~72]{R35}\\

``By definition, vacuum fluctuations occur even when there is no source
to produce particles.'' \cite[p.~124]{R35}

\section{Concluding remarks}

With the advent of the effective quantum field theory, we are now
fortunate to be aware of the existence of universal quantum fields
that fill all space all the time. These abstract fields and their
respective quanta, as listed in the Standard Model of particle physics,
constitute the ultimate ingredient of the universe disclosed to us
so far.

The idea of the underlying vista of the quantum world germinated from
Max Planck's proposal of the existence of indivisible packets of energy
called quanta. Although Planck had difficulty in believing in their
reality, Albert Einstein substantiated the veracity of quanta from
experimental observation, which led to the development of quantum
mechanics. Paul Dirac's masterful effort to combine quantum mechanics
with Einstein's special relativity eventually led to the discovery
of the quantum fields that are the ultimate ingredients of reality.

Within a decade of his proposal of the quanta, Planck also revealed
the possible existence of a zero-point energy, causing immense controversy
about their origin. With further development of quantum mechanics,
its source was identified to be thoroughly random fluctuations of
the quantum vacuum. With more advanced development, the fluctuations
are now recognized to be most likely due to spontaneous activity of
the ultimate reality of the quantum fields.\\

These facts pose at least two enigmas in a universal scale:
\begin{enumerate}
\item What causes the totally unpredictable fluctuations of the quantum
fields without involvement of a net energy? Quantum fields appear
to be nature's universal credit facility. Energy can be borrowed if
it is paid back on time. The more the amount of energy is borrowed,
the quicker it must be reimbursed. This spontaneous activity increases
tremendously at the fundamental distance scale, which corresponds
to higher energies, making the universe look like an extremely busy
place with activities having infinite randomness.
\item Yet, despite the intense activities of the quantum fields with obvious
sheer randomness, the expectation value or equivalently the average
value of the energy of the universal quantum fields have remained
immutable from almost the beginning of time. The fact that an elementary
particle of a field like an electron has exactly the same values irrespective
of when or where in the universe they are created provides conspicuous
evidence. Despite all the extreme chaos of the activity, measurement
of the value of the electron's anomalous $g$-factor gives an incredible
accuracy of one part in a trillion!
\end{enumerate}
The obvious question is: what could these enigmas possibly reveal?
Because of the lack of any other obvious option, would it be persuasive
to conclude that the incessant, spontaneous, and totally random fluctuations
of the universal quantum fields as well as the observation that the
fields remain ever immutable in spite of them, appear to be an intrinsic
property of the universe? What could it signify? Only time will tell.
Could it perhaps have something to do with Max Plank's intriguing
reply in a prominent newspaper interview, ``I regard consciousness
as fundamental. I regard matter as derivative from consciousness''
\cite{R45}?

\section{Appendix}
\label{sec:9}

The quantum harmonic oscillator has the characteristic Hamiltonian
\begin{equation}
\hat{H}= \frac{1}{2} \frac{\hat{p}^{2}}{m}+ \frac{1}{2} m\omega^{2}\hat{x}^{2} \label{eq:0}
\end{equation}
For time-independent potential energy function in the Hamiltonian,
such as $V(x)=\frac{1}{2}m\omega^{2}\hat{x}^{2}$, the time-dependent
Schr\"{o}dinger equation can be solved by separation of space $x$
and time $t$ variables \cite[pp.~25-29]{R46}, as a result of which the spatial and temporal
dependence of the ground state wave function can be factorized as
\begin{equation}
\Psi_{0}(x,t)=\psi_{0}(x)\,\exp\left(-\imath\frac{E_{0}}{\hbar}t\right)
\end{equation}
where $E_{0}=\frac{1}{2}\hbar\omega$ is the energy of the ground
state. Thus, we need to solve only the time-independent Schr\"{o}dinger
equation for the ground state position wave function
\begin{equation}
\hat{H}\psi_{0}(x)=E_{0}\psi_{0}(x)
\end{equation}
Explicitly solving the time-independent Schr\"{o}dinger equation
for the position wavefunction $\psi_{0}(x)$ of the ground state with
zero quanta, $n=0$, gives \cite[eq.~(2.60)]{R46}
\begin{equation}
\psi_{0}(x)=\left(\frac{m\omega}{\pi\hbar}\right)^{\frac{1}{4}}\exp\left(-\frac{m\omega}{2\hbar}\,x^{2}\right)\label{eq:1}
\end{equation}
The position probability density $|\psi_{0}(x)|^{2}$ can be rewritten in standard
Gaussian form as
\begin{equation}
\left|\psi_{0}(x)\right|^{2}=\frac{1}{\sqrt{2\pi\sigma_{x}^{2}}}\exp\left(-\frac{x^{2}}{2\sigma_{x}^{2}}\right)
\end{equation}
where
\begin{equation}
\sigma_{x}=\sqrt{\frac{\hbar}{2m\omega}}
\end{equation}
The momentum wavefunction of the ground state is the Fourier transform
of $\psi_{0}(x)$, namely
\begin{equation}
\psi_{0}(p)=\left(\pi m\hbar\omega\right)^{-\frac{1}{4}}\exp\left(-\frac{p^{2}}{2m\hbar\omega}\right)\label{eq:2}
\end{equation}
The momentum probability density $|\psi_{0}(p)|^{2}$ in standard Gaussian
form is
\begin{equation}
\left|\psi_{0}(p)\right|^{2}=\frac{1}{\sqrt{2\pi\sigma_{p}^{2}}}\exp\left(-\frac{p^{2}}{2\sigma_{p}^{2}}\right)
\end{equation}
where
\begin{equation}
\sigma_{p}=\sqrt{\frac{m\hbar\omega}{2}}
\end{equation}
The energy of the ground state can be computed as
\begin{equation}
E_{0}=\langle\psi_{0}(x)|\hat{H}|\psi_{0}(x)\rangle=\frac{1}{2}\hbar\omega
\end{equation}
Thus, the uncertainty relation between position and momentum follows
from the Hamiltonian of the quantum harmonic oscillator \eqref{eq:0},
which governs the temporal dynamics (quantum fluctuations) of the
ground state. It is a consequence from the quantum fluctuations of
the ground state that the position and momentum observables obey the
uncertainty relation, but it is not the other way around.

\section*{Acknowledgments}

The author wishes to acknowledge helpful discussions with Zvi Bern and Danko Georgiev.

\balance

\end{document}